\documentclass{webofc}

\usepackage[varg]{txfonts}   
\usepackage{hyperref}
\usepackage{url}
\hypersetup{colorlinks=true,citecolor=blue,urlcolor=blue,linkcolor=blue}
\begin{document}
%
\title{Pion photoproduction of nucleon excited states with Hamiltonian effective field theory }
%
%

\author{\firstname{Yu} \lastname{Zhuge}\inst{1,2,3}\fnsep
    \and
        \firstname{Dan} \lastname{Guo}\inst{4}\fnsep \and
        \firstname{Zhan-Wei} \lastname{Liu}\inst{1,2,3}\fnsep\thanks{\email{liuzhanwei@lzu.edu.cn}}
        \and
        \firstname{Derek B.} \lastname{Leinweber}\inst{5}\fnsep
        \and \\
        \firstname{Anthony W.} \lastname{Thomas}\inst{5}\fnsep 
}

\institute{School of Physical Science and Technology, Lanzhou University, Lanzhou 730000, China
    \and 
    Research Center for Hadron and CSR Physics, Lanzhou University and Institute of Modern Physics of CAS, Lanzhou 730000, China
    \and 
    Lanzhou Center for Theoretical Physics, MoE Frontiers Science Center for Rare Isotopes, Key Laboratory of Quantum Theory and Applications of MoE, Key Laboratory of Theoretical Physics of Gansu Province, Gansu Provincial Research Center for Basic Disciplines of Quantum Physics, Lanzhou University, Lanzhou 730000, China
    \and 
    School of Physics and State Key Laboratory of Nuclear Physics and Technology, Peking University, Beijing 100871, China
    \and 
   CSSM, Department of Physics, University of Adelaide, South Australia 5005, Australia
}

\abstract{Over the past few years, Hamiltonian effective field theory has been successfully applied to studies of nucleon and hyperon excited states. By discretizing the Hamiltonian in a finite volume, one can obtain the energy spectrum and compare it with the results calculated from lattice QCD. Through the analysis of experimental data, Hamiltonian effective field theory provides a framework that connects the finite-volume spectra from lattice QCD to infinite-volume scattering observables. The model independence of the approach is well preserved under the combined constraints from lattice QCD and experimental data. Building on these developments, recent works have attempted to extend HEFT to electromagnetic processes. Meanwhile, lattice QCD has also gradually advanced into the study of electromagnetic interactions. The combination of these analyses will undoubtedly deepen our understanding of light resonances.}

\maketitle
%
\section{Introduction}\label{intro}
The light nucleon excited states were discovered several decades ago. For instance, the $N^*(1440)$ resonance was first observed in 1964 by Roper et al.~\cite{Adelman:1964jmp,  Bareyre:1964pcg, Roper:1964zza}. The quark model is very successful to clarify the hadrons and explain their properties, but several problems remain unresolved, such as the so-called ``mass inversion problem'' between the $N^*(1440)$ and $N^*(1535)$ states. 

As their masses lie in the region where the nonperturbative effects of quantum chromodynamics (QCD) are significant, it is difficult to analyze these resonances directly from QCD. Chiral Perturbation Theory opened a new window for exploring the properties of such low-lying particles~\cite{Weinberg:1978kz,Veit:1984jr}. The underlying philosophy that guides our exploration of hadrons is that a physical theory serves as an effective description at a given energy scale, and accordingly, one can always construct an effective theory to study the relevant physics at that scale. Nowadays, effective field theories have been widely applied in various areas of hadron physics~\cite{Meissner:2022cbi}.

The low-lying nucleon excited states are typically excited by external probes, such as pions and photons, from the ground-state nucleons~\cite{Doring:2025wms}. These resonances couple strongly to two-particle meson–baryon channels~\cite{ParticleDataGroup:2024cfk}. They can be regarded as being dynamically generated by the coupled meson–baryon interactions, or as states that also contain contributions from an intrinsic bare core~\cite{Ramalho:2023hqd}. In partial-wave analyses, both types of contributions are usually taken into account~\cite{Matsuyama:2006rp, Briscoe:2023gmb, Drechsel:2007if, Ronchen:2014cna}. In such approaches, a phenomenological model is constructed and fitted to experimental scattering data, from which the pole positions of the resonances can be extracted. This framework has been widely employed in the study of low-lying baryon excitations.

With the advancement of computational technology, lattice QCD has enabled first-principles calculations in the nonperturbative regime. However, due to computational limitations, it is often necessary to extrapolate lattice results to the physical region with the help of  effective field theories. As a nonperturbative extension of effective field theories, Hamiltonian effective field theory (HEFT), which implements L{\"u}scher’s method~\cite{ Luscher:1986pf} provides a powerful framework for this purpose. HEFT also serves as a bridge between lattice QCD and experimental data. The coupling parameters are constrained simultaneously by lattice QCD spectra and experimental scattering data. By analyzing the components of the eigenstates, one can investigate the contributions of the bare three-quark cores and the meson–baryon channels to the physical resonances. Furthermore, HEFT has recently been extended to study the photoproduction of nucleons. 

\section{The framework of Hamiltonian effective field theory}\label{sec-2-1}

We present the case of the $S_{11}$ partial wave below. One needs to consider both the interactions between the bare states $|N^*_0\rangle$ and the two-particle channels $|\alpha\rangle$=$|\pi N\rangle$, $|\eta N\rangle$, and $|K\Lambda\rangle$~\cite{Liu:2015ktc, Abell:2023nex}:
\begin{align} G_{\alpha;N_0^*}^2(k)=\frac{3g_{\alpha;N_0^*}^2}{4\pi^2f^2}\omega_{\alpha_M}(k) u^2(k) \, , 
\end{align}
where $\alpha_M$ denotes the meson in the two-particle channel, $\omega_{\alpha_M}(k) = \sqrt{k^2 + m_{\alpha_M}^2}$, $f = 92.4~\text{MeV}$, and $u(k)$ is the form factor. In addition, one must also include the interactions among the two-particle channels themselves~\cite{Liu:2015ktc, Abell:2023nex}:
\begin{align} V_{\alpha,\beta}(k,k^{\prime})=\frac{3g_{\alpha,\beta}}{4\pi^2f^2}\frac{m_\pi+\omega_\pi(k)}{\omega_\pi(k)}\frac{m_\pi+\omega_\pi(k^{\prime})}{\omega_\pi(k^{\prime})}u(k)u(k^{\prime}) \, . 
\end{align}
The scattering amplitude can then be obtained by solving the Bethe–Salpeter equation,
\begin{align} T_{\alpha,\beta}(k,k^{\prime};E)=V_{\alpha,\beta}(k,k^{\prime})+\sum_{\gamma}\int q^2dq\frac{V_{\alpha,\gamma}(k,q)T_{\gamma,\beta}(q,k^{\prime};E)}{E-\sqrt{m_{\gamma_M}^2+q^2}-\sqrt{m_{\gamma_B}^2+q^2}+i\epsilon} \, . 
\end{align}

%
%

One can extract the $T$-matrix of $\pi N \rightarrow \pi N$, as well as the corresponding phase shifts and inelasticities. The coupling constants, $g_{\alpha;N_0^*}$ and $g_{\alpha,\beta}$, are determined by fitting to the experimental data. The fit covers the center-of-mass energy up to about 1.7~GeV, which encompasses the second $S_{11}$ nucleon excited state, $N^*(1650)$. The analysis of the $P_{11}$ partial wave can be found in Refs.~\cite{Liu:2016uzk, Wu:2017qve}. Our results can explain the experimental data for the $\pi N$ scattering well.

%

%
%
One can obtain the finite-volume potential between the bare states and the two-particle channels~\cite{Abell:2023nex}:
\begin{align} \tilde{G}_{\alpha;N_0^*}(k_n)=\sqrt{\frac{C_3(n)}{4\pi}}(\frac{2\pi}{L})^{3/2}G_{\alpha;N_0^*}(k_n) \, , 
\end{align}
where the degeneracy factor $C_3(n)$ denotes the number of integer triplets $(n_x, n_y, n_z)$ that satisfy $n = n_x^2 + n_y^2 + n_z^2$ for each integer $n$. The finite-volume 2-to-2 potential is
\begin{align}
    \tilde{V}_{i,j}(k_{n},k_{m})=\frac{\sqrt{\mathrm{C}_{3}(n)\mathrm{C}_{3}(m)}}{4\pi}\left(\frac{2\pi}{L}\right)^{3}V_{i,j}(k_{n},k_{m}) \, .
\end{align}

With the eigen-solutions of the discretized Hamiltonian, one can obtain both the mass spectrum and the composition of the eigenstates~\cite{Liu:2015ktc, Liu:2016uzk, Wu:2017qve, Abell:2023nex,Liu:2023xvy,Liu:2016wxq,Hockley:2024ipz}. The results for the $S_{11}$ spectrum can be found in Refs.~\cite{Liu:2015ktc,Abell:2023nex}. The eigenstates near the physical region, corresponding to the two low-lying odd-parity excited states, exhibit large bare-state components. The first eigenstate at light quark masses is dominated by the $\pi N$ channel. The HEFT eigenlevels can explain the mass spectra of lattice QCD, and moreover we give the reasons why some states were observed while others were not on the lattice with the components of HEFT eigenstates.

\section{Pion photoproduction off nucleons}\label{sec-2-3}
\begin{figure}[tbp]
	\centering
	\includegraphics[width=\textwidth]{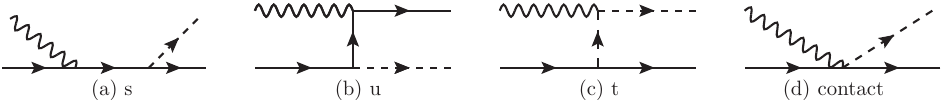}
	\caption{Tree-level diagrams for the pure electromagnetic amplitude of the $\gamma N\to \alpha$ process without FSI: (a) s channel, (b) u channel, (c) t channel, (d) the contact term. The solid, wiggly, and dashed lines represent the baryons, photons and mesons,  respectively.}\label{sut}
\end{figure}

\begin{figure}
    \centering
    \begin{tabular}{cc}
    \includegraphics[width=5.5cm,clip]{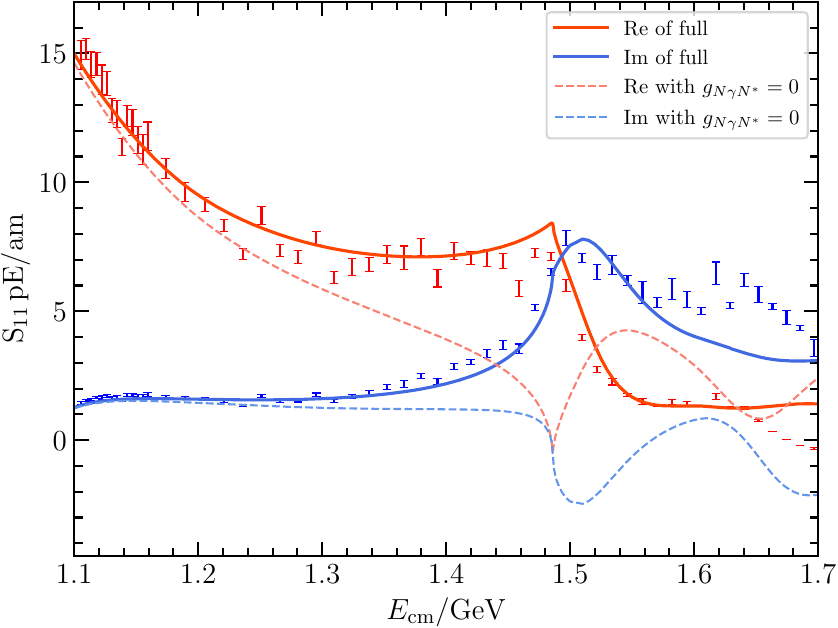} & \includegraphics[width=5.5cm,clip]{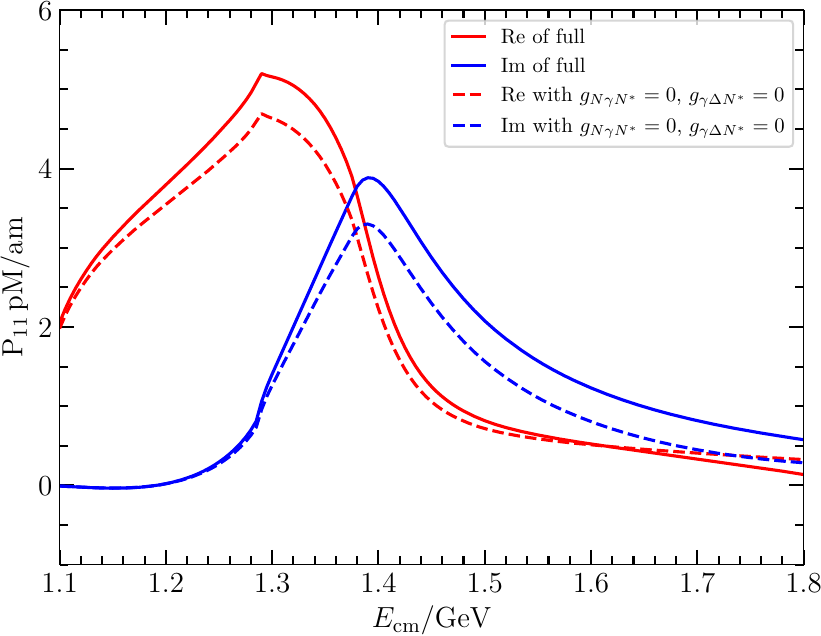}\\
    (a) & (b)
    \end{tabular}
    \caption{(a) The amplitudes $E_{0+}$ in units of attometer (am) for a proton target. (b) The amplitudes $M_{1-}$ in units of attometer (am) for a proton target. The data points are from SAID. The solid and dashed lines refer to the cases whether turning on or off the electromagnetic transition couplings between the bare nucleon resonances and nucleons, respectively.}
    \label{fig-3}
\end{figure}

With the model constrained from both experimental data and lattice QCD, we can extend HEFT to the photon induced process~\cite{Guo:2022hud, Zhuge:2024iuw}. By combining the electromagnetic transition from $\gamma N$ to the two-particle channels and the final state interactions between these channel, we can obtain the final T-matrix for $\gamma N\rightarrow\pi N$
\begin{align}
    T_{\pi N,\gamma N}^{\lambda_{\gamma},\lambda_{N}}(k,q;E)=&V_{\pi N,\gamma N}^{JLS;\lambda_{\gamma},\lambda_{N}}(k,q)+\sum_{\alpha}\int\mathrm{d}k'k^{\prime2}\frac{V_{\alpha,\gamma N}^{JLS;\lambda_{\gamma},\lambda_{N}}(k',q)T_{\pi N,\alpha}(k,k';E)}{E-\omega_{\alpha}(k')+i\epsilon} \, .
\end{align}
The multipole amplitudes can be extracted from the final $T$-matrix. In the electromagnetic potential $V_{\alpha,\gamma N}^{JLS;\lambda_{\gamma},\lambda_{N}}(k',q)$, we include the Feynman diagramsas shown in Fig. \ref{sut}. The results for the $S_{11}$ multipole $E_{0+}$ and the $P_{11}$ multipole $M_{1-}$ are shown in Fig.~\ref{fig-3}. We can see that our results are consistent with the partial wave analysis.

The coupling constants between the bare states and the $\gamma N$ channel are determined by fitting to the experimental data. If these coupling constants are set to zero, the resonances contribute only through final-state interactions. In this way, one can estimate the contributions originating from the bare states. By comparing the solid and dashed lines in Fig.~\ref{fig-3}, It is clear that the bare cores of the two $S_{11}$ resonances are indispensable for reproducing the $E_{0+}$ multipole while the absence of the bare core in the $N^*(1440)$ has only a minor effect on the $M_{1-}$ multipole. These conclusions are consistent with the HEFT analysis in the finite volume. Therefore, with the constraints from lattice QCD, as well as the experimental data from both purely hadronic channels and electromagnetic processes, one can systematically analyze whether these low-lying nucleon excited states are quark-model–like states or dynamically generated states.

%

\begin{figure}
    \centering
    \begin{tabular}{c}
    \includegraphics[width=5.5cm,clip]{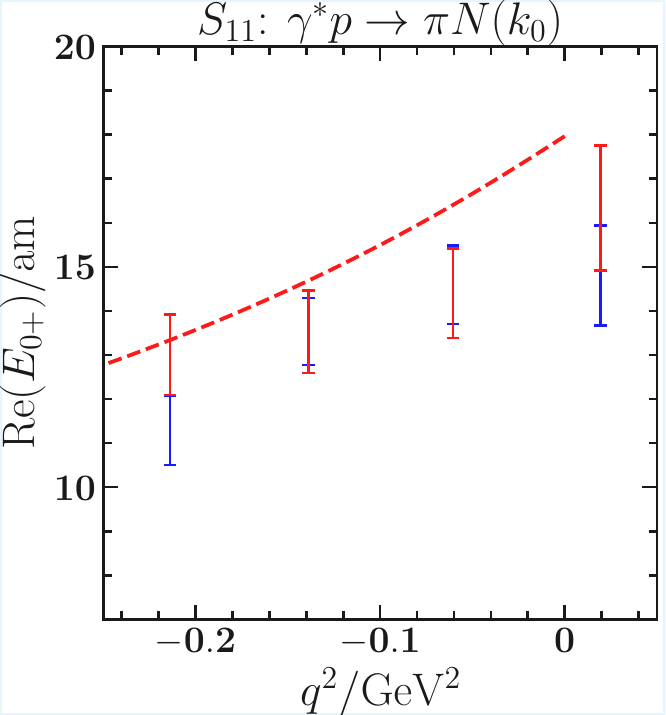}
    \end{tabular}
    \caption{The multipole amplitude $E_{0+}$ (in attometers, am) at the $\pi N$ threshold with the isospin $I_{\pi N}=1/2$. The data points are from lattice QCD simulations~\cite{Gao:2025loz}, and the red dashed line represent the extended results with HEFT.}
    \label{fig-4}
\end{figure}

The electromagnetic form factors of these nucleon excited states have also been studied using lattice QCD~\cite{Stokes:2019zdd, Stokes:2024haa}, and very recently lattice QCD simulations have for the first time investigated pion electroproduction at threshold~\cite{Gao:2025loz}. These new lattice QCD results definitely help us explore the nature of these resonances. Without adjusting the parameters in Ref. \cite{Zhuge:2024iuw}, we can easily extend the results to the case in the finite volume. The preliminary results are shown in Fig.~\ref{fig-4}, and we can see the dipole amplitudes with HEFT are consistent with the lattice QCD simulations with the finite volume effects are considered. 

\section{Summary}\label{sec-3}
The nucleon resonances play important roles in the pion photoproduction processes. Previously we have studied the $N^*(1440)$, $N^*(1535)$ and $N^*(1650)$ systematically through analyzing the $\pi N$ scattering data and lattice QCD mass spectra within the framework of HEFT~\cite{Liu:2015ktc, Abell:2023nex,Liu:2016uzk, Wu:2017qve}. The $N^*(1535)$ and $N^*(1650)$ exhibit quark-model–like structures, while $N^*(1440)$ is predominantly dynamically generated.  These features are further supported by the analyses of pion photoproduction. In the future, as lattice QCD simulations continue to advance, long-standing puzzles in hadron spectroscopy are expected to be resolved more effectively than by traditional scattering experiments alone.

\section{Acknowledgements}
This work is supported by the National Natural Science Foundation of China under Grants No. 12175091, No. 12335001, No. 12247101, the “111 Center” under Grant No. B20063, and the innovation project for young science and technology talents of Lanzhou city under Grant No. 2023-QN-107. This research is supported by the University of Adelaide and by the Australian Research Council through Discovery Projects DP210103706 (DBL) and DP230101791 (AWT).

\bibliography{ref}

\end{document}